\documentclass[prc,pdftex]{revtex4}

\usepackage{graphicx}
\usepackage{amsmath}
\begin{document}
\title{Femtoscopy in Relativistic Heavy Ion Collisions and its Relation to Bulk Properties of QCD Matter}
\author{Scott Pratt and Joshua Vredevoogd}
\affiliation{Department of Physics and Astronomy,
Michigan State University\\
East Lansing, Michigan 48824}
\date{\today}

\begin{abstract}
Using a viscous hydrodynamic model coupled to a hadronic cascade code, numerous features of the dynamics and equilibrium properties are explored for their impact on femtoscopic measurements. The equation of state, viscous parameters and initial conditions are investigated. We find that femtoscopy is affected by numerous model features at the 10\% level, and that by including features and adjusting unknown parameters, one can explain experimental source size measurements to better than 10\%.
\end{abstract}


\maketitle

\section{Overview}
The bulk properties of QCD matter, as created in relativistic heavy ion collisions, largely manifest themselves in soft hadronic observables of particles with transverse momentum less than one GeV/$c$. These observables can be divided into three classes: spectra, flow (or large-scale correlations), and correlations at small relative momentum. This last class is referred to as femtoscopy \cite{Lisa:2005dd} since these correlations are used to determine space-time characteristics of emitting sources. Correlation functions, $C({\bf P},{\bf q})$, can be linked to the outgoing phase-space distributions, or more precisely the source function $S({\bf P},{\bf r})$, which describes the probability that two particles with the same velocity, whose total momentum is ${\bf P}$, are separated by ${\bf r}$ in their asymptotic trajectory. Due to their inherent six-dimensional nature, correlations have proven to be the most difficult of all RHIC observables to fit with full dynamic models. The measurements are amenable to being fit by simple geometric models of the final state, provided that the models incorporate strong radial collective flow, and a rapid dissolution into a thermal assortment of resonances \cite{Kisiel:2008ws,Kisiel:2006is,Retiere:2003kf,Helgesson:1997zz}. However, many dynamic models, especially hybrid hydrodynamic/cascade descriptions, lead to more extended emission durations which lead to signicantly different shapes for the outgoing phase space distributions.

The information in correlations are often reduced to Gaussian source parameters, $R_{\rm out}$, $R_{\rm side}$ and $R_{\rm long}$, which are functions of the transverse momentum $k_t$, and describe the shape of the outgoing phase-space distribution of zero-rapidity particles with a specific $k_t$ and a specific azimuthal angle. Here, $R_{\rm long}$ refers to the longitudinal dimension, along the beam axis, $R_{\rm out}$ describes the outward dimension, parallel to the momentum, and $R_{\rm side}$ refers to the sideward dimension, perpendicular to the beam and to the particle's velocity. Asymptotically, the Gaussian form fits the phase space density to the form,
\begin{equation}
f({\bf p},{\bf r},t\rightarrow\infty)\sim \exp\left\{
-\frac{(r_{\rm out}-v_pt-a)^2}{2R_{\rm out}^2}
-\frac{r_{\rm side}^2}{2R_{\rm side}^2}
-\frac{r_{\rm long}^2}{2R_{\rm long}^2}
\right\}\ .
\end{equation}
The term $v_pt+a$ is irrelevant for identical particles since correlations are only sensitive to the relative position of two particles of the same velocity. For non-identical particles, one would also be sensitive to the relative position of centroids for the two species, $a_1-a_2$. 

This study focuses on how these parameters are affected by choices made in modeling the reaction. The femtoscopic relation to the equation of state has long been studied. First, a stiffer equation of state leads to more rapid expansions, with emission at earlier times and more confined to a brief burst. The reduction in the mean emission time reduces $R_{\rm long}$ as the system has less time to expand longitudinally before emission. The increased suddenness leads to a shorter $R_{\rm out}$ relative to $R_{\rm side}$, as long-lived emission allows those particles emitted earlier to move ahead of the later-emitted particles along the outward direction. Furthermore, a softer equation of state leads to higher entropy, and for fixed spectra, the entropy will be grow for increasing source volumes, $V\sim R_{\rm out}R_{\rm side}R_{\rm long}$. Since the total entropy can be ascertained in a quasi-model-independent fashion from spectra and source dimensions, and since entropy is conserved during the expansion, the product of the three dimensions is strongly linked to the equation of state \cite{Pal:2003rz}.

Femtoscopic source sizes are also affected by non-equilibrium aspects of the dynamics. Bulk viscosity, which is expected to be significant in the neighborhood of the critical temperature when the system struggles to maintain equilibrium, lowers the effective pressure and thus increases the entropy and leads to larger source dimensions. Shear viscosity is mainly important at early times, when velocity gradients are large and highly anisotropic. This leads to an enhancement in the transverse components of the energy tensor, which accelerates the transverse expansion and gives smaller values of $R_{\rm long}$ and $R_{\rm out}$ relative to $R_{\rm side}$ \cite{romatschke,Pratt:2006ss}. At the earliest times, before even viscous hydrodynamics is applicable, the pre-equilibrium state might be dominated by longitudinal color fields. These fields can, in principle, lead to exceptionally strong transverse components to the stress-energy tensor, which would amplify the effects of shear viscosity. The importance of early acceleration in explaining the experimental $R_{\rm out}/R_{\rm side}$ ratio has also been studied by incorporating initial transverse flow into ideal hydrodynamics \cite{Gyulassy:2007zz} and by adding a strong repulsive potential into microscopic cascade codes \cite{Li:2007yd}. Free-streaming during the first fm/$c$ increases the transverse pressure relative to the longitudinal pressure, which increases radial flow more than elliptic flow \cite{Heinz:2002rs}. This had been thought to make it difficult to simultaneously fit both spectra and elliptic flow, though this was accomplished in \cite{Broniowski:2008vp}. This issue might be resolved by better understanding the interface between the initial-state description and the following hydrodynamic description.

The initial density profile also affects acceleration at early times \cite{Broniowski:2008vp}. Within the typical nuclear cross-section (40 mb), a single nucleon will interact with multiple nucleons from the opposite beam. Depending on the theoretical picture, e.g., color-glass-condensate-inspired or wounded nucleon, the average radius of the initial fireball can vary by $\sim 10\%$, with a more compact source being more explosive and leading to smaller values of $R_{\rm out}/R_{\rm side}$. 

To investigate these effects we apply a relativistic viscous hydrodynamic model, which couples to a cascade code for modeling the hadronic breakup stage and the decay of outgoing resonances. The outgoing phase space points for pions are used to generate source functions, and after convoluting with the squared wave function, generate two-pion correlations. These are then treated as data and fit to correlation functions from Gaussian sources to extract $R_{\rm out}$, $R_{\rm side}$ and $R_{\rm long}$ as a function of $k_t$. The hydrodynamic code uses Israel-Stewart equations which are modified to allow one to tune to the anisotropies of the initial stress-energy tensor. Both the hydrodynamic and cascade descriptions are built on an assumption of azimuthal symmetry and boost-invariance. This prohibits a simultaneous analysis of elliptic flow, or a study of longitudinal acceleration, which is known to affect results at the 5-10\% level. In addition to investigating all the sensitivities alluded to above, we compare data from STAR collaboration at the Relativistic Heavy Ion Collider (RHIC). We do find solutions which come close to the the data without employing any particularly disquieting assumptions or any parameters outside what we would consider a reasonable range. Although this paper focuses on femtoscopy, the mean $p_t$ of various calculations is also presented and compared to data.

After reviewing details of the model in the next section, the following section is devoted to the effects of varying the equation of state, viscosities, and initial conditions. The summary is devoted to drawing conclusions with an emphasis on understanding what future improvements in models and analysis are needed to reach rigorous quantitative statements about the microscopic properties of the QCD matter formed in relativistic heavy ion collisions.

\section{The Model}
For generating interferometric source functions, phase space points are first generated from a viscous hydrodynamic model, then fed into a cascade model which models the low-density hadronic stage of the collision. Both models are written in terms of the variables $\tau$, $\eta$, $r$ and $\phi$, where $\tau=\sqrt{t^2-z^2}$ is the proper time, $\eta=\sinh^{-1}(z/\tau)$ represents the longitudinal position, and $r$ and $\phi$ represent the radial position and azimuthal angle. Both models were developed  assuming radial symmetry and boost invariance which eliminates $\eta$ and $\phi$ from consideration. By reducing the dimensionality, both speed and accuracy are vastly improved. The viscous hydrodynamic model is based on the formalism in \cite{Pratt:2007gj}, with more details provided below. The subsequent subsections provide details of the three model components.

\subsection{Viscous Hydrodynamic Model}\label{sec:hydro}

First, a review of the modified Israel-Stewart formalism described in \cite{Pratt:2007gj} is presented. A basic description of viscosity and the Navier-Stokes equation can be found in \cite{weinberg}. Recently, Israel-Stewart hydrodynamics \cite{israelstewart} has been extended and applied to nuclear physics \cite{muronga,koide,baierromatschke,heinzchaudhuri,songheinz}. In ideal hydrodynamics, the stress energy tensor becomes $P\delta_{ij}$ when viewed in the fluid rest frame (Here latin indices refer to spatial components only). Viscous hydrodynamics deals with the deviation of $T_{ij}$ from $P\delta_{ij}$. For all the hydrodynamic calculations here, the fluid rest frame is defined such that $T_{0i}=0$, and diffusion of conserved particle numbers through fluid elements is ignored. In the fluid frame the deviations of $T_{ij}$ can be expressed through five independent traceless components, $a_i$, and the deviation of the trace, $b$.
\begin{eqnarray}
\label{eq:abdef}
b&\equiv&\frac{1}{3}\left(T_{xx}+T_{yy}+T_{zz}\right)-P,\\
\nonumber
a_1&\equiv&\frac{1}{2}\left(T_{xx}-T_{yy}\right),\\
\nonumber
a_2&\equiv&\frac{1}{\sqrt{12}}\left(T_{xx}+T_{yy}-2T_{zz}\right),\\
\nonumber
a_3&\equiv&T_{xy},~~a_4\equiv T_{xz},~~a_5\equiv T_{yz}.
\end{eqnarray}
The shear components $a_i$ are related on a one-to-one basis to the five velocity gradients, $\omega_i$,
\begin{eqnarray}
\label{eq:wdef}
\omega_1&\equiv&\partial_xv_x-\partial_yv_y,\\
\nonumber
\omega_2&\equiv&\frac{1}{\sqrt{3}}\left(\partial_xv_x+\partial_yv_y-2\partial_zv_z\right),\\
\nonumber
\omega_3&\equiv&\left(\partial_xv_y+\partial_yv_x\right),~~ \omega_4\equiv\left(\partial_xv_z+\partial_zv_x\right),~~ \omega_5\equiv\left(\partial_yv_z+\partial_zv_y\right).
\end{eqnarray}
With these definitions, the Navier-Stokes equations become 
\begin{equation}
a_i=-\eta\omega_i,~~~b=-\zeta\nabla\cdot v.
\end{equation}
For Israel-Stewart equations of motion, $a_i$ and $b$ are not fixed as is the case for Navier-Stokes equations, but instead are dynamic objects. The ratios $a_i/\sigma_a$ and $b/\sigma_b$ should decay exponentially toward the Navier-Stokes values, where $\sigma_a$ and $\sigma_b$ are related to the fluctuation of the stress energy tensor at fixed energy density,
\begin{eqnarray}
\sigma_b^2&\equiv&\int d^3r \langle b(0)b({\bf r})\rangle,\\
\nonumber
\sigma_a^2&\equiv&\int d^3r \langle a_i(0)a_i({\bf r})\rangle,
\end{eqnarray}
where no sum is implied in the last expression. To restrict the values of $a_i$ and $b$ to ranges $\pm a_{\rm max}$ and $\pm b_{\rm max}$ respectively, the Israel-Stewart equations are modified by mapping $a_i$ and $b$ to $y_i$ and $x$ through hyperbolic tangents. The variables $y_i$ and $x$ will follow the Israel-Stewart equations above and can become arbitrarily large, while $a_i$ and $b$ will be restricted.
\begin{eqnarray}
\label{eq:mapping}
\frac{dy_i}{dt}&=&-\frac{1}{\tau_a}\left(y_i-\eta\omega_i/\sigma_a\right),\\
\nonumber
a&=&a_{\rm max}\tanh\left(\frac{\sigma_ay}{a_{\rm max}}\right),~~y=\sqrt{y_1^2+y_2^2},\\
\nonumber
a_i&=&a\frac{y_i}{y},~~~a=\sqrt{a_1^2+a_2^2},\\
\nonumber
\frac{dx}{dt}&=&-\frac{1}{\tau_b}\left(x-\zeta\nabla\cdot v/\sigma_b\right),\\
\nonumber
b&=&b_{\rm max}\tanh\left(\frac{\sigma_bx}{b_{\rm max}}\right).
\end{eqnarray}
As derived in \cite{Pratt:2007gj} and \cite{jou}, the lifetimes, fluctuations and viscosities are not independent,
\begin{equation}
\label{eq:tausigmavisc}
\eta=\frac{\sigma_a^2\tau_a}{T}, ~~~\zeta=\frac{\sigma_b^2\tau_b}{T}.
\end{equation}

The equations of motion are solved by storing the velocities and energy densities in a mesh defined by the radial coordinate. Following the ideas of the one-dimensional calculation in \cite{Pratt:2008jj}, mesh points are not stored at equal times but at varying times that enforce local simultaneity, i.e., $u\cdot\Delta x=0$, where $\Delta x$ is the four vector describing the separation of two neighboring mesh points. Using the integrated distance along the mesh as seen by comovers,
\begin{equation}
\ell= \int d\ell, ~d\ell=\sqrt{-[dx-u(u\cdot dx)]^2},
\end{equation}
the acceleration in the fluid frame, which equals the rate of change of the  transverse rapidity, takes on a simple form,
\begin{equation}
\frac{d}{d\tau}y_i=a(\ell)=-\frac{\partial_\ell T_{xx}}{T_{00}+T_{xx}}.
\end{equation}
In order to maintain simultaneity between neighboring mesh points, the time step depends on $\ell$,
\begin{equation}
\delta\tau(\ell)=\delta\tau(\ell=0)\exp\left\{\int_0^\ell d\ell'~a(\ell')\right\}.
\end{equation}
To complete the equations of motion, an expression is needed for the evolution of the energy density. This is done by considering the change in the internal energy within a cell defined by adjacent mesh points and a fixed small rapidity range $\delta\eta$. In the fluid frame, the volume of the cell is
\begin{equation}
\Delta V=(\tau\delta\eta) 2\pi R\Delta\ell,
\end{equation}
where $R$ is the radius as viewed in the laboratory frame. In a time step $d\tau$ the volume increases both due the increase in the longitudinal dimension $d(\tau\delta\eta)$ and by the increase in the transverse dimensions. Writing
\begin{eqnarray}
d\Delta V&=&d\Delta V_x +d\Delta V_z,\\
\nonumber
d\Delta V_x&\equiv&(\tau\delta\eta)d(2\pi R\Delta\ell),~~~d\Delta V_z=(2\pi R\Delta\ell)d(\tau\delta\eta),
\end{eqnarray}
the change in the internal energy of the cell is
\begin{equation}
d\Delta U=-T_{xx}d\Delta V_x-T_{zz}d\Delta V_z,
\end{equation}
where $z$ is the longitudinal direction and $x$ refers to the radial direction. Given the internal energy and volume, one then knows the local energy density $\epsilon$ which closes the equations of motion. For the ideal case, $T_{ij}=P\delta_{ij}$, one recovers $dU=-PdV$, which implies entropy conservation. Indeed, when the code was run in this limit, entropy was conserved to better than 0.2 percent.

The equation of state used for the runs shown here consisted of three parts. For temperatures below 170 MeV, the equation of state was that of a hadronic gas. For a given cell, the pressure was calculated as a function of the energy density and the density of five conserved charges. The conserved charges were the number of strange plus anti-strange quarks, the number of baryons plus antibaryons, the effective number of pions (e.g., a $\rho$ meson counts as two pions), the number of $\eta$s and the number of $\omega$s. Only the standard octet mesons and octet and decuplet baryons were considered, i.e., the $\pi, K, \eta, \rho,\omega,K^*, \eta',\omega,\phi$ mesons and the $p,n,\Lambda,\Sigma,\Xi,\Delta,\Sigma^*,\Xi^*,\Omega$ baryons. The details of which particle numbers were fixed was not particularly important because the breakup density was chosen to be 400 MeV/fm$^3$, which allowed the cells very little time to adjust their chemistry before the evolution was taken over by the cascade code. 

For an intermediate range of energy densities, $\epsilon_h<\epsilon< \epsilon_h+L$, the equation of state was chosen to have a constant speed of sound, i.e., $P=P_h+c^2_{\rm mixed}(\epsilon-\epsilon_h)$. Here, $\epsilon_h$ is the energy density of an equilibrated hadron gas with a temperature of 170 MeV. In the limit of $c_{\rm mixed}=0$, the equation of state becomes that of a first order phase transition with latent heat $L$. For energy densities above $\epsilon_h+L$, the speed of sound was bumped up to $c^2=0.3$ to be consistent with lattice gauge theory \cite{Cheng:2007jq}. The simple form for the equation of state was used so that by varying $L$ and $c_{\rm mixed}$ one could study the sensitivity to the equation of state.

The ratio of the shear viscosity to entropy was fixed above $T_c$. According to the KSS conjecture, this ratio should stay above $1/4\pi$ \cite{kss}. Results for varying $\eta/s$ are investigated in Sec. \ref{sec:sensitivities}. The fluctuation $\sigma^2_a$ was calculated by considering fluctuations of a massless gas. This gives $\sigma_a^2=(4\pi/5)T^2s$. The relaxation time from Eq. (\ref{eq:tausigmavisc}) is then $\tau_a=(5/4\pi)\eta/Ts$. For $\epsilon<\epsilon_h$, the relaxation time was chosen as $\tau_a=1/(n\sigma)$, with $\sigma=25$ mb and $n$ being the hadron density. This energy-averaged cross-section for a hadronic gas would be somewhat higher than 25 mb, but much of that cross section would be more forward peaked which reduces the effectiveness of collisions to thermalize the matter. It would not be surprising if more sophisticated calculations of the relaxation time would differ by a few tens of percent. The fluctuation $\sigma_a$ was determined by considering the fluctuations inside a hadron gas,
\begin{equation}
\sigma_a^2=\int d^3r~T_{xy}(0)T_{xy}(r)=\frac{1}{V}\sum_{{\rm particles}~i} \frac{p_{i,x}^2p_{i,y}^2}{E_i^2}
=\sum_{{\rm species}~\alpha}(2j_\alpha+1)\int \frac{d^3p}{(2\pi)^3} f_\alpha(p) \frac{p_x^2p_y^2}{E^2}.
\end{equation}
The relaxation time is then given by Eq. (\ref{eq:tausigmavisc}). For the intermediate region, $\epsilon_h<\epsilon<\epsilon_h+L$, both $\eta$ and $\sigma_a^2$ were chosen to vary linearly with the energy density from the hadronic value at $\epsilon_h$ to the value for the lower end of the plasma region. Relaxation times were then chosen according to Eq. (\ref{eq:tausigmavisc}).

Bulk viscosities are expected to be negligible except near $T_c$ \cite{Paech:2006st}. This can be understood by considering an isotropically expanding thermalized gas, i.e., a Hubble expansion, which for photons maintains a thermal form to the photon spectrum with the temperature falling $\sim 1/\tau$. For a non-relativistic gas, a thermal form also ensues, but with the temperature falling as $1/\tau^2$. If the shape of the momentum-space distribution is already thermal, collisions are needed to maintain the thermal value for $T_{ii}$. Bulk viscosities thus disappear high above $T_c$ where the temperatures far exceed the quark mass, and for temperatures well below the pion mass. 

In contrast, near $T_c$ the degrees of freedom and the condensed fields need to change in order to maintain equilibrium, which leads to a bulk viscosity in that region \cite{Paech:2006st,Karsch:2007jc}. The bulk viscosity was chosen to be zero for $\epsilon>\epsilon_h+L$ and for $\epsilon<\epsilon_h$. In the middle of the mixed region, $\epsilon=\epsilon_h+L/2$, $\zeta$ was set to a maximum value, $\zeta_{\rm max}$. To make the bulk viscosity a continuous function, it was chosen to linearly fall with energy density above and below the maximum value so that it returned to zero at the boundary of the mixed region. Arbitrarily, the relaxation time was chosen to be $5\hbar/(4\pi T)$, which equals the minimum relaxation time for the shear in the plasma phase if one is at the KSS limit. The treatment of the effects of bulk viscosity here are undoubtedly naive. Since the principal source of bulk viscosity might be the non-equilibrium chiral condensate, or $\sigma$ field, relaxation times might be very large, the response might be very non-linear, and the behavior might be oscillatory, which contradicts the Israel-Stewart assumption that non-equilibrium deviations decay exponentially. Thus, the investigations can probably only qualitatively describe the impact of non-equilibrium effects towards the trace of the stress-energy tensor. Bulk effects due to non-equilibrium fields would be better treated by a simultaneous solution of the respective wave equations coupled to the hydrodynamic medium.

\begin{figure}
\centerline{\includegraphics[width=0.5\textwidth]{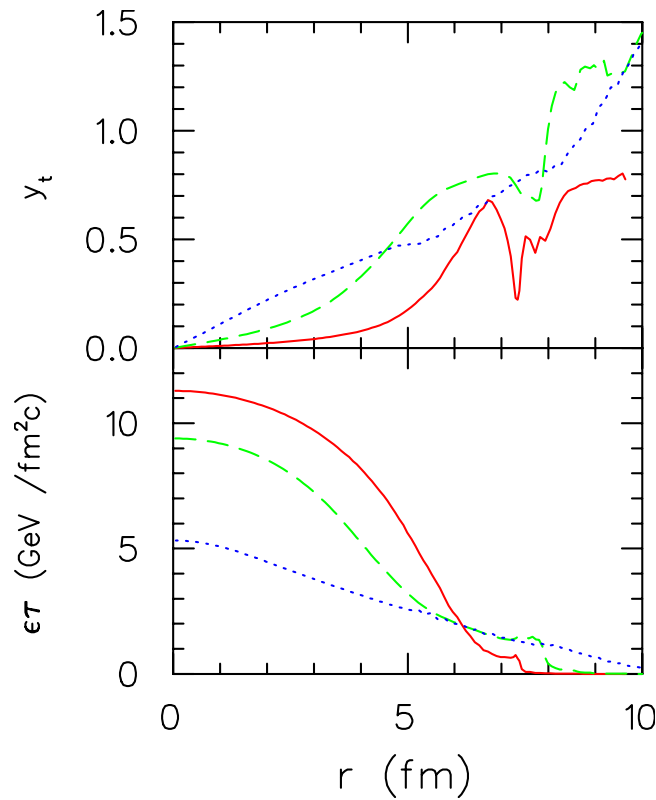}}
\caption{\label{fig:profile}(color online)
The transverse rapidity (upper panel) and energy density (lower panel, multiplied by $\tau$) profiles for three times: $\tau=1$ fm/$c$ (solid line), $\tau=3$ fm/$c$ (dashed line) and $\tau=6$ fm/$c$ (dotted line). The lower speed of sound near $T_c$ causes a tsunami-like pulse to grow which then largely dissipates before breakup. 
}
\end{figure}

\begin{figure}
\centerline{\includegraphics[width=0.5\textwidth]{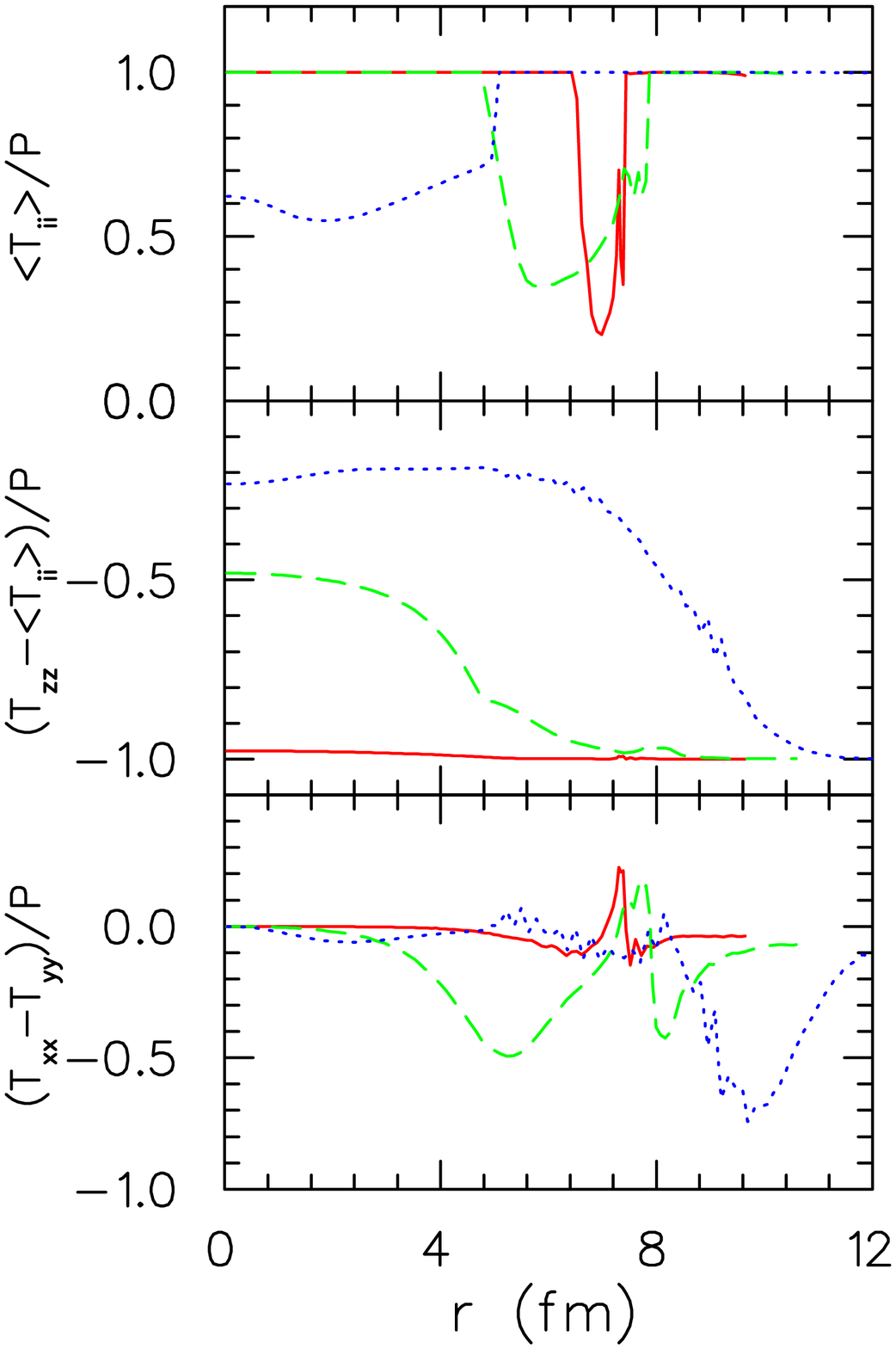}}
\caption{\label{fig:stressenergy}(color online)
Deviations of the stress-energy tensor due to viscous effects are displayed as a function of $r$ for three times:
$\tau=1$ fm/$c$ (solid line), $\tau=3$ fm/$c$ (dashed line) and $\tau=6$ fm/$c$ (dotted line). The effective pressure, $\langle T_{ii}\rangle\equiv (T_{xx}+T_{yy}+T_{zz})/3$ scaled by the pressure (upper panel) deviates from unity due to bulk viscosity which is only non-zero near $T_c$. The longitudinal components $(T_{zz}-\langle T_{ii}\rangle)/P$ begins at the saturation value of -1 enforce by Eq. (\ref{eq:mapping}). The ratio moves toward zero as the velocity gradient lessens, but less so for large $r$ due to the large viscosity at low density. The lower panel shows that an anisotropy in $(T_{xx}-T_{yy})/P$ grows mainly in the region where the outgoing radial pulse creates a pulse in the radial component of the velocity gradient.
}
\end{figure}

Figures \ref{fig:profile} and \ref{fig:stressenergy} display the hydrodynamic evolution for the default parameter set. The kinks in the collective transverse rapidity shown in Fig. \ref{fig:profile} arise from the region near $T_c$. As the matter expands into a region with lower speed of sound a pulse builds up similar to a tsunami. Bulk viscosity, which lowers the effective pressure, $\langle T_{ii}\rangle\equiv(1/3)(T_{xx}+T_{yy}+T_{zz})$, in this region amplifies the pulse. The pulse largely dissipates by the time final breakup occurs as the rapidity profile becomes linear and the energy density profile also becomes smooth.

Viscous effects on the stress energy tensor are illustrated in the three panels of Fig. \ref{fig:stressenergy}. The effective pressure $\langle T_{ii}\rangle$ is displayed in the upper panel. Since the bulk viscosity is set to zero outside the intermediate region, the ratio $\langle T_{ii}\rangle/P$ varies from unity only in this region. If not for the saturation enforced by Eq. (\ref{eq:mapping}), the effective pressure might fall below zero. The size of the effect is enhanced by the pulse which results in large velocity gradients at the boundaries of the pulse. The anisotropy of $T_{ij}$ is shown in the lower two panels. At $\tau_0$ the anisotropy was inserted as a boundary condition with $T_{zz}$ set to zero, which equivalently gives $(T_{zz}-\langle T_{ii}\rangle)/P=-1$ as shown in the middle panel. This is also the saturated value as enforced by Eq. (\ref{eq:mapping}), and is maintained for some time due to the large velocity gradients at early times. At the edge of the fireball, where the matter is in the lower density hadronic phase, the strong anisotropy remains due to the large viscosity at low density. However, the behavior in this region is somewhat irrelevant as it is below the breakup density and is handled by the cascade description described in the next section. The lower panel shows $T_{xx}-T_{yy}$, which differs from zero mainly near $T_c$ due to the radial pulse. The large variations shown in Fig. \ref{fig:stressenergy} shows how Navier-Stokes treatments, which can lead to arbitrarily large deviations of the stress-energy tensor, are questionable at early times or in the region of the radial pulse. 


The hydrodynamic module was run until the entire system fell below the breakup density. A sampling of emitted hadrons was generated from the entire evolution. At each time step, particles were generated from thermal surface emission of the outermost cell whose energy density was above the breakup density. The generation of particles was consistent with the temperature, density, and the anisotropy of the stress-energy tensor. When a cell's energy density fell below the breakup density, particles were emitted from that cell in the same manner, except according to volume emission. In order to make such emission consistent with the anisotropy of the stress energy tensor, the particles were first generated according to an isotropic thermal distribution. The momenta $p_x$, $p_y$ and $p_z$ as seen in the fluid frame were then scaled by factors $\lambda_x$, $\lambda_y$ and $\lambda_z$ respectively, where 
\begin{eqnarray}
\lambda_i=\sqrt{T_{ii}/P}.
\end{eqnarray}
The same mechanism is used for surface emission, along with an additional factor taking into account the rate at which the particles leave the surface. For a given particle moving along a collision-less trajectory, the momenta as measured in the local rest frame should scale inversely with the time between collisions. For each component of the momentum, $p_{\rm local}=p_0\tau_0/(\tau+\tau_c)$, where $\tau_0$ is the inverse velocity gradient at the time of the last collision along the given direction and $\tau_c$ is the time since the last collision. For non-relativistic particles one can derive the simple scaling form for the various momenta shown above. However, for lighter particles the simple scaling is only approximately justified. Future versions of the program will apply a more sophisticated mechanism for generating particles. 

\subsection{Hadronic cascade Model}\label{sec:cascade}
For energy densities below 400 MeV/fm$^3$, a cascade code is used to describe the evolution. The cascade simulates the evolution of the particles as straight line trajectories, punctuated by collisions whose probability is determined by a combination of a fixed cross section of 15 mb, along with resonant absorptions and decays. The resonant cross sections use a simple Breit-Wigner form with fixed lifetimes, and all collisions and decays are treated as $s$-waves. Only resonances from the standard meson octets or from the baryon octet and decuplet are included. This is a simple treatment, with no mean field or Bose effects, but should provide a sufficiently reasonable description of the breakup stage for the interferometric studies presented here.

Particles are entered into the cascade description from a Monte Carlo list generated by the hydrodynamic module. Along with the list of particles, the cascade module is also given a description of the position of the emitting surface as a function of time. Any particle that returns to the interior of the surface during the cascade description is deleted. Assuming that the hydrodynamic code, with its inclusion of shear anisotropies, accurately models the behavior of a hadron gas for energies near the 400 MeV/fm$^3$, this should provide a fully consistent interface. For all of the parameter sets studied here, the hyper-surface from which particles are created by the hydrodynamic module rapidly collapses resulting in time-like emission for the vast majority of particles. The percentage of particles that are re-absorbed into the hyper-surface during the cascade is only about one percent. 

One potential issue with many cascade codes is that the finite interaction range, i.e., particles collide at a finite interaction range of approximately $\sqrt{\sigma/\pi}$, leads to viscous effects \cite{Cheng:2001dz}. Usually, such effects are minimized by oversampling the distribution by a factor $N_{\rm sample}$ which reduces the cross sections by $1/N_{\rm sample}$ and the interaction ranges by $1/\sqrt{N_{\rm sample}}$. However, in the cascade description applied here, the interaction range is set to zero, thus eliminating such numerical viscosities. This is accomplished by exploiting boost invariance and azimuthal invariance, which allows the trajectories to be treated as radii evolving as a function of the proper time, $r(\tau)$. When two particles have the same radius, a probability is calculated for their colliding given the cross section and the fact that the sampling covers $2\pi$ radians and one unit of $\eta$. Given that the two radii are equal and that the other coordinates are irrelevant given the symmetries, the effective interaction range is zero. Furthermore, since any correlations from collisions or resonant decays are spread out over a wide range of $\eta$ and $\phi$, this treatment should come extremely close to a true Boltzmann description even though particles are represented on a one-to-one basis.

The algorithm is optimized by storing the information for each particle in a list ordered by radius. A second ordered list stores the list of pairs that will cross and is ordered by crossing times. The crossings are executed in order of time, and since the list is ordered, new crossings need only be calculated for the nearest neighbors of those particles that have crossed. When two particles cross, one needs to calculate the probability that they collide, or merge to form a resonance. This is related to the ratio of the cross section to the area of the cylinder over which the particles are spread, $2\pi r\tau$ (assuming one unit of $\eta$ is being modeled). The exact probability is complicated by the relative angles of the particles and relativistic effects, and is given by:
\begin{eqnarray}
{\rm Scattering~Probability}&=&\frac{\sigma}{2\pi R\tau}\frac{\sqrt{-q^2P^2}}{P_0q_x-P_xq_0},\\
\nonumber
P^\alpha&=&p_1^\alpha+p_2^\alpha,~~q^\alpha=(p_1-p_2)^\alpha-P^\alpha\frac{P\cdot(p_1-p_2)}{P^2}.
\end{eqnarray}
Here $x$ refers to the radial components.

\begin{figure}
\centerline{\includegraphics[width=0.5\textwidth]{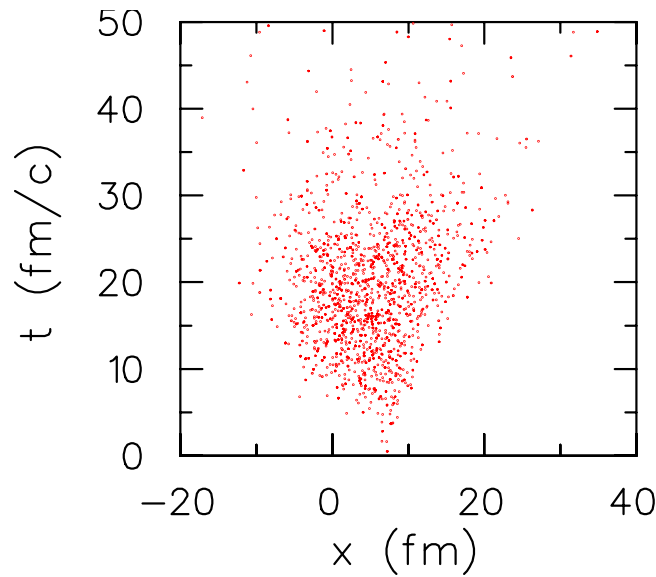}}
\caption{\label{fig:xyzt}(color online)
Positions of last interaction for pions with $p_x=300$, $p_y=0$ MeV/$c$. The outward position has a modestly positive correlation with emission time. This correlation is indicative of an exploding source. Even though the duration emission is rather long, over 10 fm/$c$, the correlation allows $R_{\rm out}/R_{\rm side}$ to be close to be close to unity. 
}\end{figure}

All collisions before 25 fm/$c$ are simulated, with decays being performed until they are exhausted. Weak decays are allowed to take place, except for charged kaons, pions and $K_{\rm long}$ mesons. The point at which each particle had its last collision is recorded to be used for calculating spectra and femtoscopic correlations. A sampling of phase space points is displayed in Fig. \ref{fig:xyzt} for particles with transverse momentum $p_t=300$ MeV/$c$, and reveals a modest positive correlation between the outward position and time. This is opposite what one expects for an inwardly burning source whose emitting surface would move inward with time. Instead, it suggests an exploding source. 

A glance at Fig. \ref{fig:xyzt} reveals a strikingly different source than that extracted from Blast-wave models. First, both the average lifetime and the duration of the emission are longer than those determined by blast-wave models \cite{Retiere:2003kf,Kisiel:2008ws}. Secondly, whereas the blast-wave analysis of \cite{Kisiel:2008ws} suggested a negative $x-t$ correlation, there is a modest positive correlation in Fig. \ref{fig:xyzt}. Such a positive correlation was also seen in the AMPT model \cite{ampt}, which is based on a cascade picture for both partons and mesons. Since the hyper-surface representing the transition from hydrodynamics to the Boltzmann approach has a negative $x-t$ correlation, this emphasizes the importance of accurately accounting for the breakup dynamics with a microscopic model. An underestimate of the emission duration for blast-wave models is expected if the blast-wave model employed an $x-t$ correlation of the wrong sign, as a positive correlation allows one to maintain a small $R_{\rm out}/R_{\rm side}$ ratio despite a longer emission duration. The mean emission time is associated with the ratio $R_{\rm long}/v_{z,{\rm therm}}$, where $v_{z,{\rm therm}}$ is the thermal velocity for longitudinal motion. Since blast wave analyses typically ignore shear effects and thus assume thermal motion is locally isotropic, they would overestimate $v_{z,{\rm therm}}$ if in fact the local momentum distribution is broader in the transverse plane than along the longitudinal direction. An overestimate of $v_{z,{\rm therm}}$ would lead to an underestimate of the lifetime.

The algorithm is both efficient and accurate. The procedure eliminates artifacts associated with particles interacting at a finite separation, and running on a single CPU, an event is performed in less than ten seconds. However, the approach has one numerical disadvantage. When calculating the value of $\tau=\sqrt{t^2-z^2}$ at which two particles will reach the same radius, one must solve quartic equations. Numerical errors for such solutions are non-negligible which occasionally lead to particles being propagated in such a way that violates the ordering by radius. This calculation is performed millions of times within a single event, and the violations tend to occur approximately once per every ten events. In such an instance, the event is abandoned. This does not seriously detract from the numerical efficiency, but such failures significantly complicated the construction of the code, as frequent error checking is required. 

\subsection{Generating and Fitting Correlation Functions}\label{subsec:cf}

Correlations can be generated via the Koonin equation,
\begin{eqnarray}
\label{eq:koonin}
C({\bf P},{\bf q})&=&\int d^3r~S({\bf P}/2,{\bf r})\left|\phi({\bf q},{\bf r})\right|^2,\\
\nonumber
S({\bf P},{\bf r})&=&\frac{\int d^4x_ad^4x_b~ s({\bf P}/2,x_a)s({\bf P/2},x_b)\delta({\bf r}-({\bf x}'_a-{\bf x}'_b))}
{\int d^4x_a d^4x_b~s({\bf P}/2,x_a)s({\bf P/2},x_b)},
\end{eqnarray}
where ${\bf P}$ is the pair's momentum, ${\bf r}$ is the spatial separation of the particles in the frame of the pair, and $\phi$ is the outgoing relative wave function. The probability of emitting a pion of momentum ${\bf p}$ from space-time point $x$ is $s({\bf p},{\bf x})$, with $x'$ being the coordinate in the pair frame. The source function $S({\bf P},{\bf r})$ is simply the normalized probability that two particles of the same momentum, ${\bf P}/2$, are separated by ${\bf r}$ in the pair's center of mass. The relative wave function incorporates quantum symmetrization, and both the Coulomb and strong interaction between the two pions. For the calculations shown here ${\bf q}$ refers to one half the relative momentum as measured in the pair frame.

In practice, the Koonin equation is straight-forward to implement. To calculate $C({\bf P},{\bf q})$, one first extracts the subset of phase-space points from the output of the Boltzmann codes whose transverse momenta are within 5 MeV/$c$ of ${\bf P}/2$. For every pair in the subset, one calculates $|\phi({\bf q},{\bf r})|^2$ for an array of ${\bf q}$ values. The same set of pairs is used for every value of ${\bf q}$. The correlation function is then the average of $|\phi|^2$ for the pairs. Statistics for such calculations are greatly enhanced by the rotational and boost invariances, as every particle's phase space points can be rotated and boosted so that it has zero longitudinal momentum and travels in a given azimuthal direction.

\begin{figure}
\centerline{\includegraphics[width=0.5\textwidth]{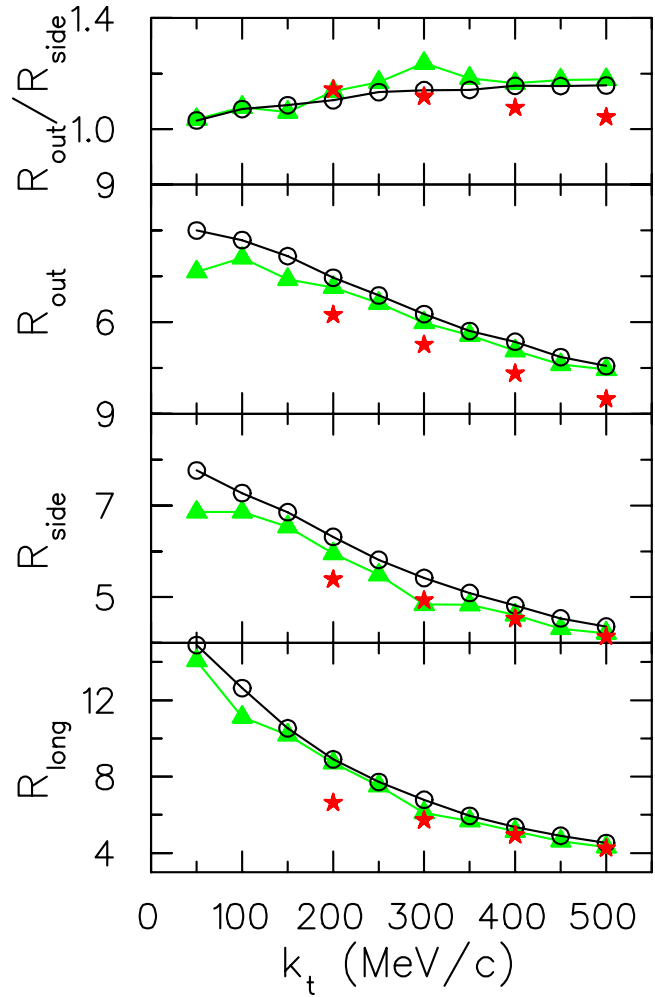}}
\caption{
\label{fig:Rosl_fullwf}(color online)
Gaussian source sizes were found by fitting generated correlation functions to those from Gaussian sources. For the default parameter set, correlations were calculated from Eq. (\ref{eq:koonin}) with (triangles) and without (circles) the effects of Coulomb and strong interaction in the relative wave function. Since calculations are much quicker without the interactions, and since differences are small, interactions are neglected for the calculations in the next section. The differences between the default calculation differs from experimental values (stars) \cite{starhbt} are significantly lessened as compared to calculations based on ideal hydrodynamics \cite{Soff:2000eh,Heinz:2004et}.}
\end{figure}

If one neglects the inter-pair Coulomb and strong interactions, the calculations can be greatly accelerated by calculating
\begin{equation}
\rho({\bf q})=\sum_i e^{2i{\bf q}\cdot{\bf r}_i},
\end{equation}
where the sum covers the $N$ particles in the subset used above, and $r_i$ is the position of the $i^{\rm th}$ particle. For large $N$, correlations can be generated by simply squaring the sum,
\begin{equation}
C({\bf q})=1+\frac{1}{N^2}|\rho({\bf q})|^2.
\end{equation}
Since one never has to evaluate a double-sum, this method is quicker than the alternate method. Although it can only be applied if one neglects strong and Coulomb forces, this method should be sufficient if one is generating correlations for the purpose of finding effective Gaussian source sizes. Both methods were tried for the calculation with the default parameters, with the comparison being illustrated in Fig. \ref{fig:Rosl_fullwf}. Since the differences in the extracted Gaussian source sizes were small and the trends of interest are unlikely to be affected, the latter method was chosen. Although the calculations using the full wave functions are more realistic, it should be pointed out that experimental analyses have generated source radii by dividing the experimental correlation function by a $q-$dependent factor with the purpose of dividing away the effect of the Coulomb force in affecting correlation funcitons \cite{starhbt,phenixhbt}. Since the Coulomb correction factors are based on isotropic Gaussian sources, the procedure is not exact. Now that the discrepancy between models and experiments are less than 10\%, the errors introduced in this procedure should be re-examined. In particular, errors should be checked for radii at higher $k_t$. For $k_t\sim 500$ MeV/$c$, source functions are highly anisotropic in the frame of the pair due to Lorentz contraction, with $R_{\rm out}$ approaching five times $R_{\rm side}$, whereas the correction factors are built assuming that the source shape is isotropic in that frame.

To generate Gaussian radii, correlations were calculated on a three-dimensional mesh in ${\bf q}$. These were then compared to predictions for $C({\bf q})$ for Gaussian sources. The radii were chosen to minimize the sum of the squared radii. Even though the correlations were remarkably non-Gaussian for $q<10$ MeV/$c$, the radii were remarkably robust, and did not change appreciably if one neglected the low $q$ points in the fit. Since the mesh and $q$ values were generated in the pair frame, the outward source size was then Lorentz contracted so that it represented the shape of the outgoing phase space density in the laboratory frame.

\section{Femtoscopic Ramifications of Adjusting the Equation of State, Viscosity and Initial Conditions}
\label{sec:sensitivities}

One of the prime motivations for interferometric measurements was the possibility of observing a long-lived mixed phase, which could only happen if the equation of state was first order with a large latent heat. Some bag model parameterizations employed in the early days of the field had latent heats of several GeV/fm$^3$. If that were the case, extracted lifetimes at the AGS and SPS might have been several tens of fm/$c$ depending on the amount of initial stopping \cite{Rischke:1996em,Pratt:1986cc}. The longest lifetimes would occur for conditions where the interior energy density was initially at the peak value for the mixed phase. Since a mixed phase has zero sound velocity, there would no impetus for explosion, and instead the outside would emit hadrons like a burning log. Lattice calculations now preclude such equations of state, and indeed, no such long lived phases have been observed. Instead, in lattice calculations the speed of sound appears to dip down to about $c_s^2\sim 0.1$, before re-stiffening to $\sim 0.3$ at high temperatures \cite{Cheng:2007jq}. By including resonances in a hadron gas, the speed of sound is expected to be $\sim 0.15$ below $T_c$. Given that initial energy densities at RHIC are well above those of the soft region, one should expect RHIC collisions to be more explosive and shorter-lived that those at the AGS and SPS. Qualitatively, these expectations have been met. However, quantitatively describing the source sizes with full hydrodynamic models has proved elusive. Femtoscopy provides a six-dimensional test of any dynamical model, so it should not be surprising that reproducing experimental source sizes requires using a realistic equation of state, accurately modeling viscous effects and using correct initial conditions. We explore the impact of each of these three aspects of the modeling in the next three subsections. Results from the default parameter set are compared to results where an isolated parameter set has been adjusted. For each calculation, the initial energy density is adjusted so that the final $dN_{\rm ch}/d\eta\sim 690$ \cite{phobosdndeta}.

Radial flow, and in turn spectra, are also affected by all the variations studied here. Table \ref{table:meanpt} presents the mean $p_t$ for pions, kaons and protons. Again, it should be emphasized that these calculations include all weak decays of hyperons and of the $K_s$. To some extent, these decay products are subtracted from experimental analyses, which might lead to the model predictions under predicting the mean $p_t$ for pions. However, the calculations also neglect symmetrization effects on the pion spectra, which should lower the pions mean $p_t$. Unfortunately, the uncertainties in the experimental values in Table \ref{table:meanpt} are rather large, mainly due to the fact that experiments measure in a finite $p_t$ range. The experimental values for mean $p_t$ do agree with the default calculation, within the large experimental uncertainties. A more meaningful comparison, which would involve comparing actual spectra in the measured regions, is outside the scope of this study, but Table \ref{table:meanpt} is certainly sufficient for evaluating the sensitivity of the spectra to the various model parameters studied here. It should be emphasized that the sensitivity of spectra to the equation of state has been considered by numerous authors, and that in \cite{Huovinen:2005gy}, sensitivity of the elliptic flow is also considered.

\begin{table}
\begin{tabular}{|r|c|c|c|}\hline
	&$\pi^{(+,0,-)}$	&$K^{(+,-)}$	&$p,n,\bar{p},\bar{n}$\\ \hline\hline
STAR\cite{starpt}& $422\pm 22$	& $719\pm 74$	& $1100\pm 110$	\\ \hline
PHENIX\cite{phenixpt}& $453\pm 33$	& $674\pm 78$	& $954\pm 85$	\\ \hline\hline
L=0	& 528	& 897	& 1310	\\ \hline
L=800 MeV/fm$^3$	& 433	& 714 	& 1027	\\ \hline
L=1.6 GeV/fm$^3$	& 403	& 652	& 931 	\\ \hline\hline
$c_s^2=0$	& 406	& 659	& 945	\\	\hline
$c_s^2=0.1$	& 433	& 714 	& 1027	\\	\hline
$c_s^2=0.2$	& 463	& 772	& 1116	\\	\hline\hline
$4\pi\eta/s$=0	& 408	& 664	& 957	\\ \hline
$4\pi\eta/s$=2	& 433	& 714 	& 1027	\\ \hline
$4\pi\eta/s$=4	& 449	& 743	& 1081	\\ \hline
Initially isotropic & 428 & 695 & 1012	\\ \hline\hline
$4\pi(\zeta/s)_{\rm max}=0$	& 462	& 763	& 1107	\\ \hline
$4\pi(\zeta/s)_{\rm max}=2$	& 433	& 714 	& 1027	\\ \hline
$4\pi(\zeta/s)_{\rm max}=4$	& 418	& 679	& 983	\\ \hline\hline
CGC IC	& 447  	& 741	& 1062	\\ \hline
Wounded Nucleon	& 433	& 714 	& 1027 \\ \hline
Collision Scaling	& 482	& 806	& 1173	\\ \hline
\end{tabular}
\caption{\label{table:meanpt}
The mean $\langle p_t\rangle$ in MeV/$c$ for central collisions for pions, kaons and protons. Only charged species were used in the PHENIX analysis, and only negative hadrons were used for STAR. Increasing the stiffness of the equation of state or the shear viscosity raises $\langle p_t\rangle$ for heavier particles due to the corresponding increase in radial flow. The increase of radial flow with shear viscosity derives from the increase in $T_{xx}$ and $T_{yy}$ relative to $T_{zz}$ at early times. By beginning the calculation with an isotropic stress energy tensor, radial flow is modestly reduced. Increasing the bulk viscosity lowers the effective pressure, and thus modestly reduces radial flow. Using collisional scaling to set the initial energy density profile results in a more compact initial source, which then generates more radial flow.}
\end{table}

\subsection{Adjusting the Equation of State}
\begin{figure}
\centerline{\includegraphics[width=0.5\textwidth]{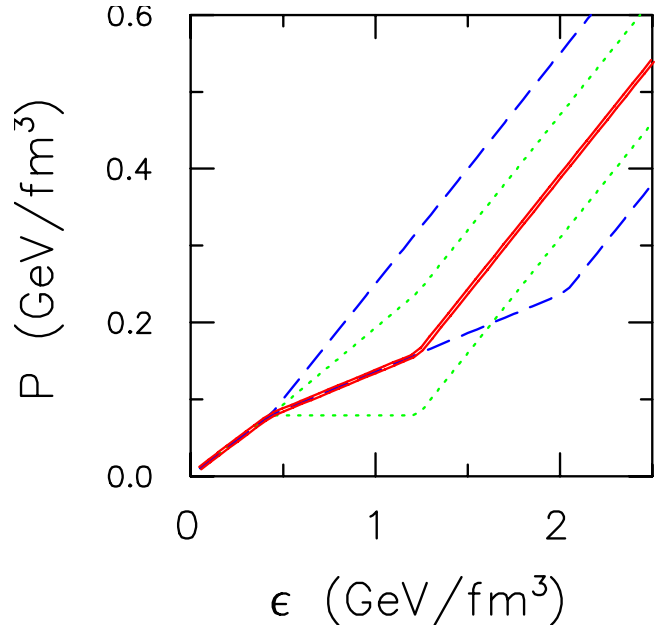}}
\caption{\label{fig:eqofst}(color online)
Pressure vs. energy density for five equations of state: The default equation of state (solid line) assumed a constant speed of sound, $c_s^2=0.1$, for an intermediate range of energy densities, $\epsilon_h<\epsilon<\epsilon_h+L$, where $L=800$ MeV/fm$^3$. The speed of sound in the intermediate region was varied (dotted lines) to either 0.2 or zero. The latter choice corresponds to a first order phase transition. Keeping the default speed of sound, the width of the intermediate region was also varied (dashed lines) to either zero or 1.6 GeV/fm$^3$.}
\end{figure}

To study the sensitivity to the equation of state, we vary both the speed of sound and the width of the intermediate region, with the five equations of state being displayed in Fig. \ref{fig:eqofst}. For temperatures below 170 MeV, or equivalently for energy densities below $\epsilon_h\sim 400$ MeV/fm$^3$, the equation of state is that of a resonance gas. For the intermediate region, $\epsilon_h<\epsilon<\epsilon_h+L$, the equation of state has a constant speed of sound, $P-P_h=c_s^2(\epsilon-\epsilon_h)$. Above the intermediate region, the speed of sound is set to $c_s^2=0.3$, to be consistent with lattice calculations at high temperature.

\begin{figure}
\centerline{\includegraphics[width=0.5\textwidth]{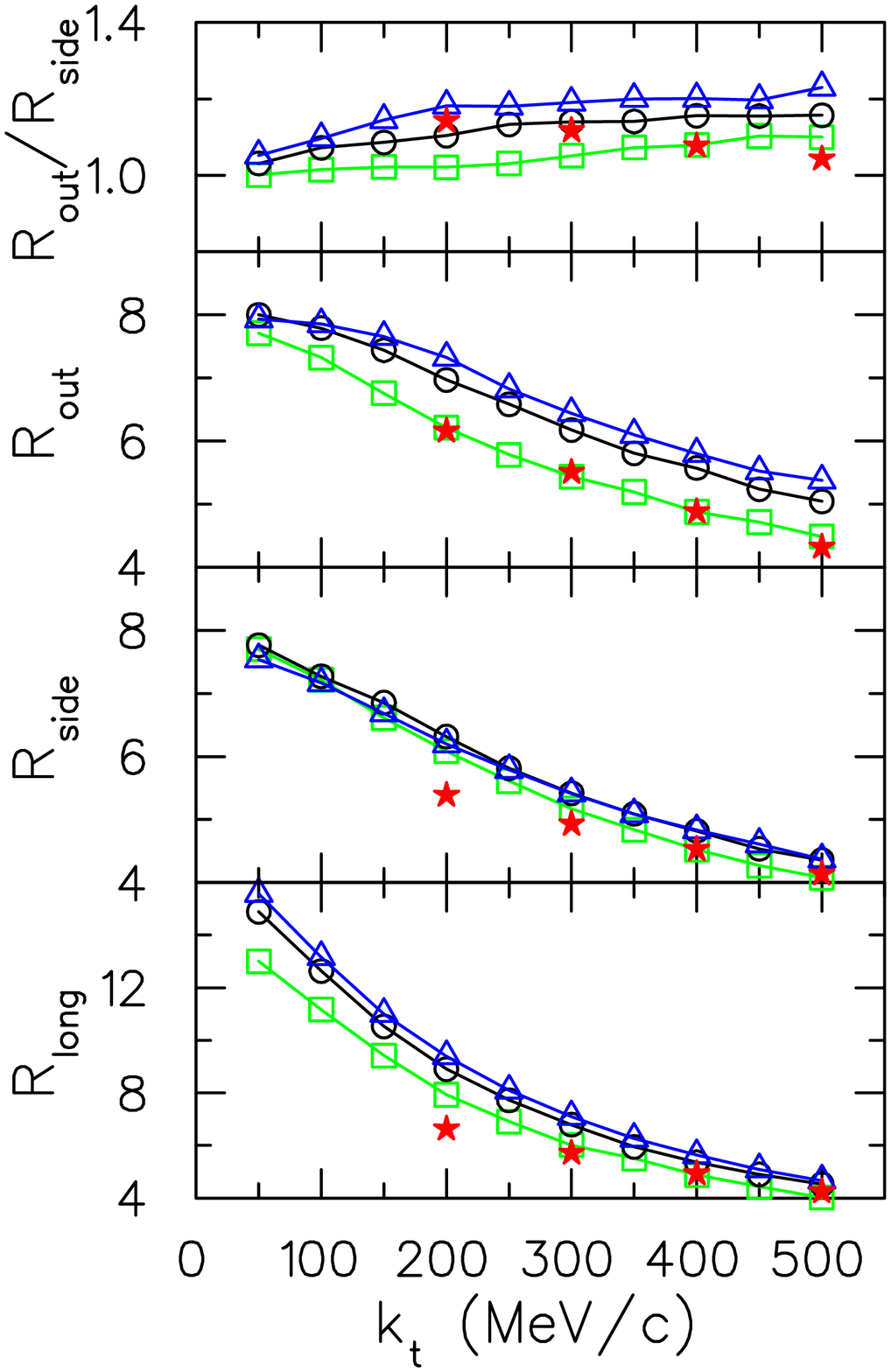}}
\caption{\label{fig:Rosl_L} (color online)
Gaussian source dimensions for different equations of state. The equation of state incorporated a soft region where the speed of sound was set to $c_s^2=0.1$. The width of that region was $L=800$ Mev/fm$^3$ in the default calculation (circles) and is compared to $L=0$ (squares) and $L=1.6$ GeV/fm$^3$ (triangles). The softer equations of state yield larger values of $R_{\rm out}/R_{\rm long}$. Experimental values are also depicted (stars).
}
\end{figure}
\begin{figure}
\centerline{\includegraphics[width=0.5\textwidth]{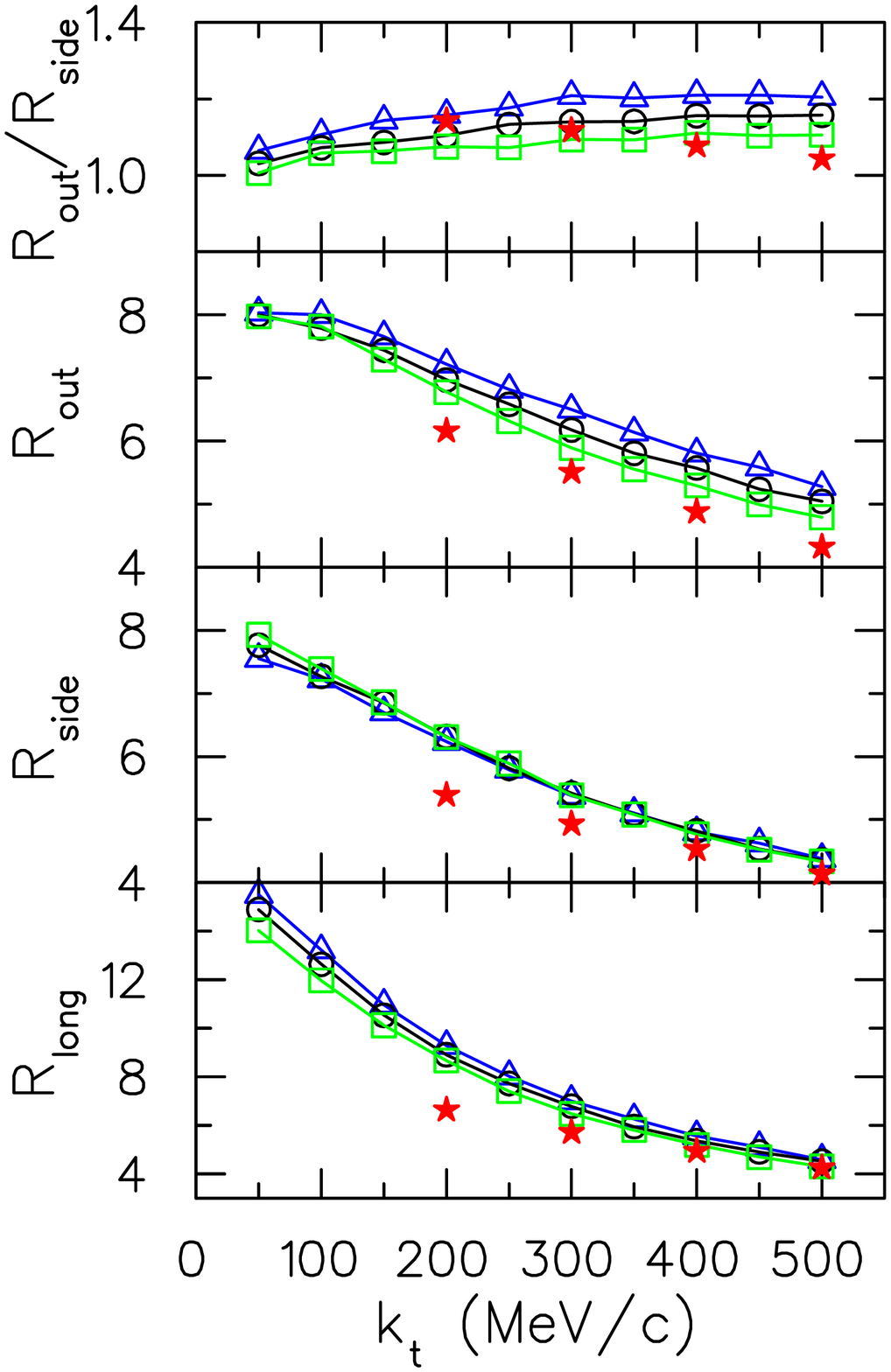}}
\caption{\label{fig:Rosl_c2}(color online)
Gaussian source dimensions for different equations of state. Rather than varying the width of the soft region as was shown in Fig. \ref{fig:Rosl_L}, the speed of sound in the soft region was varied from the default value $c_s^2=0.1$ (circles) to $c_s^2=0$ (triangles) and $c_s^2=0.2$ (squares). Experimental values are also depicted (stars).
}
\end{figure}

The equation of state can be softened by either increasing $L$ or decreasing the mixed-phase value of $c_s^2$. Figure \ref{fig:Rosl_L} displays source sizes for the three values of $L$: the default value of 800 MeV/fm$^3$, a soft value of 1.6 GeV/fm$^3$, and a hard value of zero. The default value was chosen to be crudely consistent with the behavior of lattice calculations which show a strong stiffening of the matter for energy densities rising from $1-1.5$ GeV/fm$^3$. The equation of state was also altered by adjusting the mixed-phase value of $c_s^2$ from the default value of 0.1 to either a stiffer value of 0.2 or to a softer value of zero, which would correspond to a first-order phase transition. The femtoscopic effect of varying the speed of sound is shown in Fig. \ref{fig:Rosl_c2}.

As expected, softer equations of state lead to longer relative values of $R_{\rm long}$ and $R_{\rm out}$ relative to $R_{\rm side}$. Whereas the increase in $R_{\rm long}$ signals an increase in the mean emission time, the increase in $R_{\rm out}$ is indicative of a longer duration of the emission, or a more outside-in nature to the emission. The product of the three radii increases for softer equations of state. This is due to the increase in entropy associated with a softer equation of state (when compared at the same energy density). Although these variations in the equation of state are rather strong, doubling $L$ or $c_s^2$, femtoscopic radii were affected on the level of 10\%. Spectra are also affected by changes to the equation of state at the level of 10\% as seen in Table \ref{table:meanpt}. In particular softer equations of state lower collective radial flow which leads to lower values of the mean $p_t$, especially for heavier particles.

\subsection{Adjusting Viscosities}\label{sec:varyvisc}

Even modest viscosities signficantly modify the stress energy tensor. It has been proposed that shear viscosity can not fall below the KSS limit, $\eta\ge s/4\pi$ \cite{kss}. According to Navier Stokes, at early times where the velocity gradient is $1/\tau$, the KSS limit yields
\begin{eqnarray}
T_{zz}&=&P-\frac{s}{3\pi\tau},~\sim P\left(1-\frac{4}{3\pi T\tau}\right),\\
\nonumber
T_{xx}&=&P+\frac{s}{6\pi\tau},~\sim P\left(1+\frac{2}{3\pi T\tau}\right),
\end{eqnarray}
where the expressions involving $P$ assumed a free gas equation of state, $P=\epsilon/3$. One expects for thermalization times near 1/2 fm/$c$ where $T\tau\sim 1$, that the correction to the longitudinal pressure is $\sim 40\%$. If the viscosity is more than twice the KSS bound the value of $T_{zz}$ can become negative. One expects a higher shear to accelerate the radial flow and result in lower values of $R_{\rm long}$ and $R_{\rm out}/R_{\rm side}$ \cite{Paech:2006st}, as well as increased $\langle p_t\rangle$ for heavier particles. These expectations have already been demonstrated by Romatschke \cite{romatschke}.

The default calculation presented here assumes that the shear viscosity is twice the KSS bound. This would yield negative $T_{zz}$ at early times, if not for the mapping described in Eq. (\ref{eq:mapping}) which restricts the modification from shear to be less than the absolute value of the pressure. In the default calculation the initial condition for the stress energy tensor was set to this maximum value, with $T_{zz}=0$ and $T_{xx}=T_{yy}=\epsilon/2$, consistent with color-glass calculations \cite{Krasnitz:2002mn}. For pure non-interacting classical fields, the anisotropy would be even larger as $T_{zz}$ is negative, $T_{zz}=-\epsilon$ and $T_{xx}=T_{yy}=\epsilon$. 

\begin{figure}
\centerline{\includegraphics[width=0.5\textwidth]{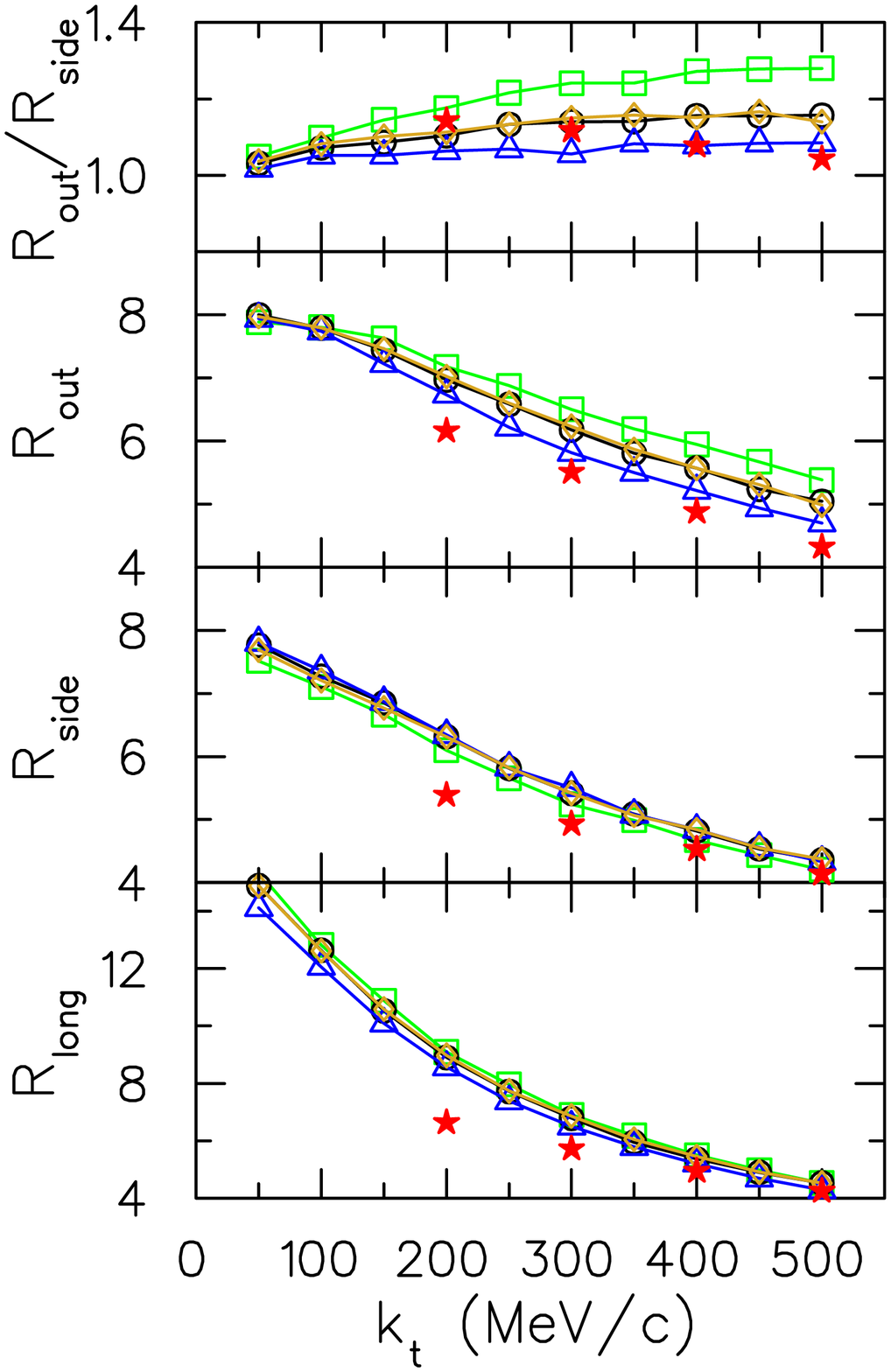}}
\caption{\label{fig:Rosl_visc}(color online)
Gaussian source dimensions for three different shear viscosities in the high-energy phase: Results from the default calculation with $4\pi\eta/s=2$ (circles) are compared to results using $4\pi\eta/s=0$ (squares) and $4\pi\eta/s=4$ (triangles). Higher shear viscosities result in more rapid initial accelerations and smaller $R_{\rm out}/R_{\rm side}$ ratios. Also shown is a modification of the default calculation where the stress-energy tensor was initialized as isotropic, rather than at the saturated values. Even though non-negligible variances resulted for the source radii, differences were found for the mean $p_t$ seen in Table \ref{table:meanpt}. Experimental values are also depicted (stars).
}
\end{figure}
In addition to the default calculation, three modifications to the shear viscosity are illustrated here. First, the viscosity in the high density phase is set to zero. For the first modification the initial anisotropy to the stress energy tensor is also set to zero. For the second modification, the viscosity in the high-density phase is doubled relative to the default calculation to four times the KSS bound. Due to the ceiling imposed on the viscous modifications, the higher viscosity only matters for times greater than 1 fm/$c$. For the final modification, the default calculation is modified by setting the initial anisotropy of the stress energy tensor to zero. This mainly affects the expansion during the first one fm/$c$. Since the Israel-Stewart relaxation times tend to be $\sim 1/2$ fm/$c$, memory of the initial condition is lost after that point. 

The expectation for the femtoscopic radii are borne out by the results illustrated in Fig. \ref{fig:Rosl_visc}. The default calculation differs from the zero-viscosity calculation by $\sim 10$\%. In particular, the $R_{\rm out}/R_{\rm side}$ ratio comes significantly closer to unity. It should be pointed out that the zero-viscosity calculation differs from many previous ideal hydrodynamic calculations in that the initial time was set to 0.1 fm/$c$, whereas several other calculations used either 0.6 or 1.0 fm/$c$, which would further increase the $R_{\rm out}/R_{\rm side}$ ratio. However, there is no physical justification for setting $T_{xx}=T_{yy}=0$ at early times, thus such an initial state seems unwarranted. This is discussed in more detail in \cite{Vredevoogd:2008id}. Modifying the initial anisotropy is similar in principal to altering the viscosity for early times.

Bulk viscosity is only expected to be significant near the critical region. In particular, the condensed fields may not be able to keep pace with rapidly changing equilibrium values. This can lead to a peak in the bulk viscosity in the intermediate energy region \cite{Paech:2006st}, which has been verified with analysis of lattice results \cite{Karsch:2007jc}. The divergence of the velocity $\nabla\cdot v$, incorporates velocity gradients in all three directions is approximately one third (fm/$c$)$^{-1}$ for $\tau=5$ fm/$c$. For the default calculation, the peak value the $\zeta$ in the intermediate region is $2s/4\pi$. The magnitude of the effect is similar to what was derived in \cite{Paech:2006st}. For such a velocity gradient the trace of the stress-energy tensor is modified by a substantial fraction. Doubling the bulk viscosity can make the Navier-Stokes value of $\langle T_{ii}\rangle< 0$. For the calculations performed here, the mapping procedure of Eq. (\ref{eq:mapping}) saturates the size of the change in $\langle T_{ii}\rangle$ to be less than $P$. However, when combined with shear effects individual components can fall below zero. It should be emphasized that the non-equilibrium effects that generate bulk viscosity, mainly non-equilibrium fields, may be very poorly represented by viscous formalisms. First, in the transition region, responses may be highly non-linear, and secondly the field might not relax exponentially toward equilibrium as is assumed in Israel-Stewart treatments. Thus, the study here can really only point at the qualitative impact of bulk viscosity on the dynamics, and ultimately on the femtoscopy. 

\begin{figure}
\centerline{\includegraphics[width=0.5\textwidth]{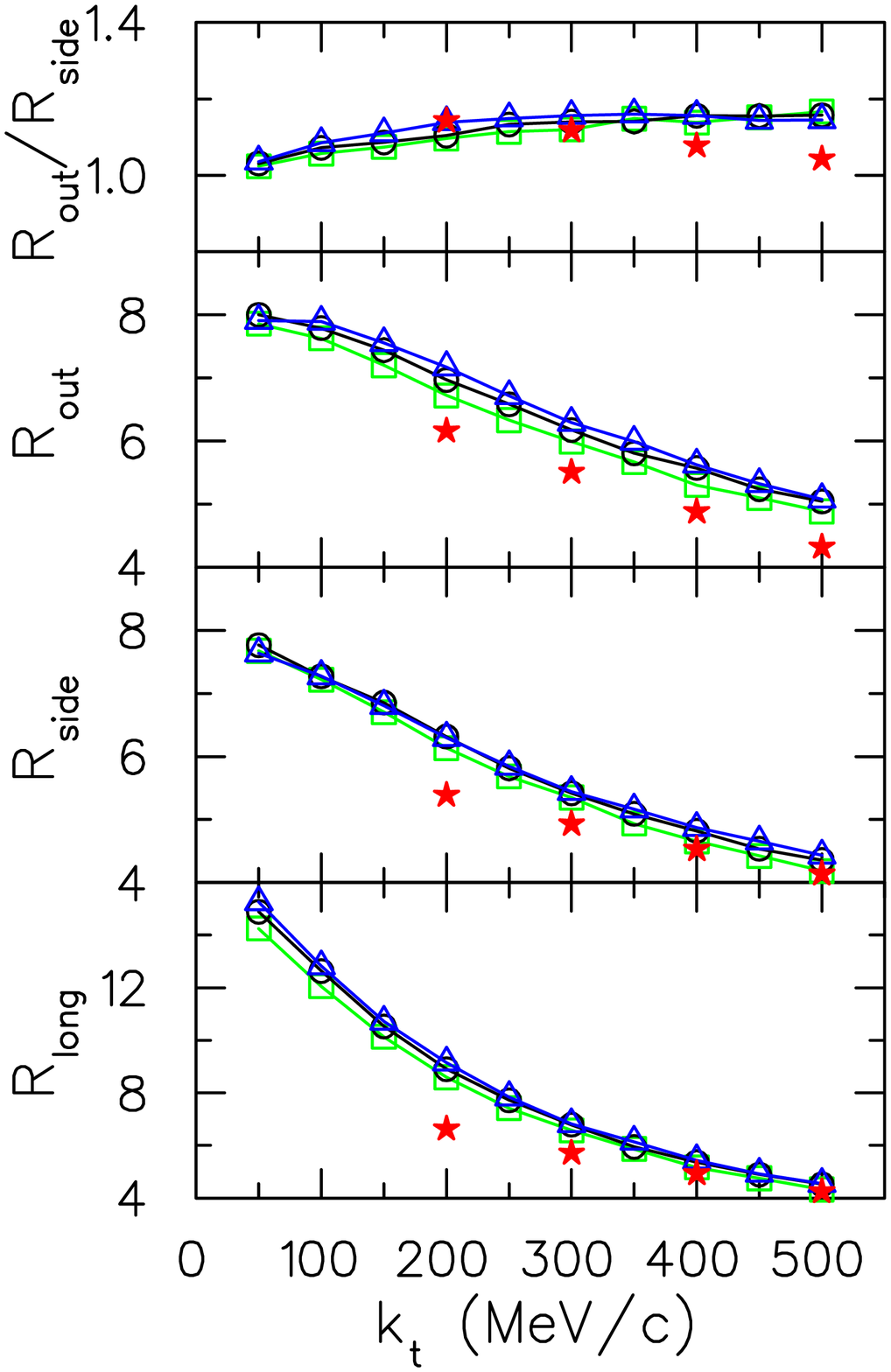}}
\caption{\label{fig:Rosl_B}(color online)
Gaussian source dimensions for three different bulk viscosities. Bulk viscosities were set to zero outside the intermediate-energy density region near $T_c$, and varied linearly from zero to a maximum value in the center of the region. The three values, $4\pi \zeta_{\rm max}/s$= 0 (squares), the default value 2 (circles) and 4 (triangles), resulted in nearly identical radii. Somewhat larger differences were observed in the mean $p_t$ values from Table \ref{table:meanpt} and in the initial energy densities from Table \ref{table:epsilon0}. Experimental values are also depicted (stars).
}
\end{figure}
The default calculation was modified twice, once by doubling the bulk viscosity and once be eliminating it. The difference of the three calculations, illustrated in Fig. \ref{fig:Rosl_B}, was modest. The effect on the stress energy tensor is similar to what one would get by softening the equation of state in the transition region. But unlike the changes to the equation of state investigated in the previous sub-section, these changes do not affect the pressure at either high or low density. Hence, the impact on observables tended to be modest. Increasing the bulk viscosity only affected the femtoscopy at the level of a few percent. The bulk viscosity helped amplify the magnitude of the pulse in the energy density created near the soft region. However, the pulse largely dissipated later in the collision. The bulk viscosity had a slightly more significant impact on the mean $p_t$ as seen in Table \ref{table:meanpt} and in the initial energy density, which was adjusted to match $dN/d\eta$, as seen in Table \ref{table:epsilon0}.

Bulk viscosity had a visible impact on the smoothness of the energy density profiles. Larger bulk viscosities appeared to lead to jagged and unstable profiles in the intermediate region. We speculate that this is driven by the fact that bulk viscosity effectively pushes the pressure vs. energy density to behave non-monotonically, which could give regions where $d\langle T_{ii}\rangle/d\epsilon$, which is the effective speed of sound squared, is zero or negative. It would be interesting to know whether such instabilities would appear in a more physically grounded treatment of non-equilibrium effects, such as one where the dynamics of non-equilibrium fields were treated in parallel to the hydrodynamic treatment by solving a coupled Klein-Gordon equation.

The significant sensitivity of the final femtoscopic source sizes to acceleration during the first one or two fm/$c$ might seem surprising. The importance of early acceleration can be likened to an olympic sprint, where a head start of a few tenths of a second results in a difference of several meters at the end of the race.  For this reason, it is imperative to understand the bulk properties, e.g., the stress-energy tensor, of matter even before thermalization.

\subsection{Adjusting Initial Conditions}
For a rotationally and boost invariant calculation the choice of initial conditions involves choosing the initial transverse density, the initial stress energy tensor, and the initial time. For all calculations presented here, the initial time was chosen to be 0.1 fm/$c$. Since the development of collective flow at such early times is driven by the initial stress energy, there is no reason to pick a later time, unless there were reason to expect $T_{xx}$ and $T_{yy}$ to be zero at early times. Since a little more than 0.1 fm/$c$ is required for the nuclei to pass one another, and given that this is already an extremely short time relative to the overall expansion time, there is no motivation to pick an earlier time. Variations of the initial stress energy tensor, and in particular variations to the initial anisotropy of $T_{ij}$ were considered in the previous subsection along with variations in the viscosity. 

Three variations of the  initial energy density profile were explored. The default calculation was that of the wounded nucleon model \cite{woundednucleon}. In this calculation the probability of a nucleon interacting is calculated as unity minus the probability it survives without interaction. The survival probability is calculated assuming the particle travels through the Woods-Saxon nuclear profile with a 40 mb cross section. For the thick part of the nucleus, this approaches participant number scaling. An alternative is collision scaling, where the energy density at a transverse coordinate $(x,y)$ is proportional to $T_aT_b$, where the thickness function $T$ is calculated by integrating the density of a nucleus over the $z$ coordinate. The third profile explored here is for the color-glass profile used in \cite{Drescher:2007cd}. In that case the energy density is chosen proportional to the minimum of $T_a$ and $T_b$. For all three profiles, the thickness functions were found by convoluting two Woods-Saxon profile whose centers differed by an impact parameter of 2.21 fm, corresponding to the 5\% most central collisions of Au+Au. The density profile was then averaged over the azimuthal angle to generate an approximate radial profile. For every calculation, the profiles were renormalized so that the resulting $dN_{\rm ch}/d\eta$ was $691\pm 5$, consistent with \cite{phobosdndeta}.

\begin{figure}
\centerline{\includegraphics[width=0.5\textwidth]{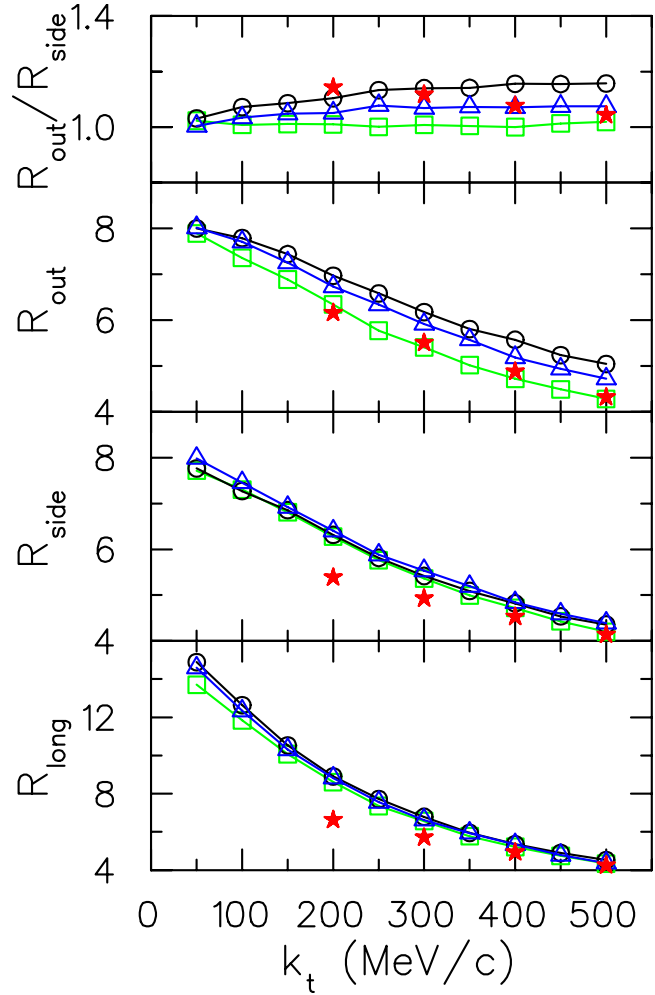}}
\caption{\label{fig:Rosl_IC}
(color online) Gaussian source dimensions for three different initial energy density profiles: the wounded nucleon model (circles) is used for the default calculation, and results in a less compact source than either the color-glass inspired model of \cite{Drescher:2007cd} or collision scaling. More compact sources are more explosive and lead to lower $R_{\rm out}/R_{\rm side}$ ratios. Experimental values are also depicted (stars).
}
\end{figure}

The collision-scaling profile was the most compact of the three attempted here, and resulted in the largest radial flow. This profile resulted in higher $\langle p_t\rangle$ for protons and lower values of $R_{\rm out}/R_{\rm side}$ as seen in Fig. \ref{fig:Rosl_IC}. The least compact profile was the default calculation which was based on the wounded nucleon model. 

The initial energy density at the extreme periphery of the collision should be driven by collisional scaling since one can ignore the possibility of multiple collisions in that limit. None of the three profiles employed here obey this constraint, as the overall normalization is scaled in order to match the experimental value $dN_{\rm ch}/d\eta$. If this constaint were replaced by the constraint that the density profile behaved correctly for small $T_a$ and $T_b$, the experimental $dN_{\rm ch}/d\eta$ could instead additionally constrain the remaining parameters. The experimental $dN_{\rm ch}/d\eta$ has been used to argue that the profile appears more driven by participant scaling than collision scaling \cite{phobosdndeta}. However, these analyses are exceedingly simple and neglect many aspect of the expansion such as viscosity or longitudinal work. Since the temperature is not constant across the profile, the relation between entropy and energy densities is not linear, thus different profiles are reached if one believes the energy density should follow a given scaling vs. the entropy density, which is more related to the multiplicity. This underscores the importance of performing a global analysis of all observables, including elliptic flow, which is also affected by the choice of profile shape \cite{Drescher:2007cd}.

It is unfortunate that the initial energy density is not itself an observable. After adjusting the initial energy density for each parameter set to match the experimental $dN_{\rm ch}/d\eta$, the final state observables tend to change at the 10\% level or less for all variations studied here. However, the initial energy densities vary by more than a factor of two for these calculations as seen in Table \ref{table:epsilon0}. This variation has little to do with the variation of the asymptotic transverse energy for fixed multiplicity, which is also a $\sim 10$\% effect. The variation largely derives from differences in the longitudinal work in the expansion. The work is proportional to both $T_{zz}$ and the time over which the system expands. For large viscosities, softer equations of state, or for initial conditions where $T_{zz}$ is small, the longitudinal work is reduced, thus requiring smaller initial energy densities to attain a given final condition. For shear anisotropies $T_{xx}$ and $T_{yy}$ are also enhanced, which accelerates the expansion. This results in a reduced time for the reaction, which also reduces the longitudinal work. Changing the shape of the initial profile also changes the energy density, mainly for the trivial reason that a more compact initial profile requires a higher energy density to produce the same $dN_{\rm ch}/d\eta$.

\begin{table}
\begin{tabular}{|c|c|c|c|}\hline
width of soft region in EoS & $\epsilon_0$ (GeV/fm$^3$)\\ \hline
L=0	&	150\\ \hline
L=800 MeV*	&	114.5\\ \hline
L=1.6 GeV	&	104.5\\ \hline\hline
stiffness of soft region in EoS & \\ \hline
$c_s^2=0$	&	107\\	\hline
$c_s^2=0.1$*	&	114.5\\	\hline
$c_s^2=0.2$	&	124.5\\	\hline\hline
shear viscosity in parton phase & \\ \hline
$4\pi\eta/s$=0	&	289\\ \hline
$4\pi\eta/s$=2*	&	114.5\\ \hline
$4\pi\eta/s$=4	&	106.5\\ \hline
initially isotropic init. cond. &	148\\ \hline\hline
max. bulk viscosity in soft region & \\ \hline
$4\pi \zeta_{\rm max}/s$=0	&	124 \\ \hline
$4\pi \zeta_{\rm max}/s$=2*	&	114.5\\ \hline
$4\pi \zeta_{\rm max}/s$=4	&	109\\ \hline\hline
initial density profile & \\ \hline
CGC IC	&	136\\ \hline
Wounded Nucleon*&	114.5\\ \hline
Collision Scaling	&	180\\ \hline
\end{tabular}
\caption{\label{table:epsilon0}
Initial central energy density at $\tau_0=0.1$ fm/$c$, in GeV/fm$^3$. Values, which were adjusted to match experimental values of $dN_{\rm ch}/d\eta$, vary by more than a factor of two, largely due to differences in longitudinal work. The default calculation (noted by *) is varied in five ways.}
\end{table}

\section{Summary and Outlook}
The principal conclusion from these investigations is that the femtoscopic data from RHIC can be reproduced to within 10\% with models combining viscous hydrodynamics and hadronic cascades. In particular, the $R_{\rm out}/R_{\rm side}$ ratio can be brought down close to unity. The failure of previous models appears to derive mainly from three shortcomings, all of which are related to under-predicting the explosivity of the collision. First, the equations of state were often too soft, using a first-order phase transition. A stiffer equation of state is more explosive, and can lower the $R_{\rm out}/R_{\rm side}$ ratio. Secondly, previous treatments ignored acceleration before the thermalization time. From general arguments involving conservation of energy and momentum in the equations of motion of the stress-energy tensor, it should be clear that thermalization is not required for acceleration. In fact, longitudinal classical fields, which are far from equilibrium by definition, result in strong transverse acceleration. Finally, the previous treatments were based on ideal hydrodynamics. The effects of shear, as already demonstrated in \cite{romatschke}, increase the transverse pressure relative to the longitudinal pressure at early times, which of all the variations considered here, appears to be the most important. Bulk effects, were manifest in the final mean $p_t$, but made remarkably little difference in femtoscopic radii. Previous treatments overpredicted the $R_{\rm out}/R_{\rm side}$ ratio by 40\% or more \cite{Soff:2000eh}, a result confirmed if we run this model with a softer equation of state, without viscosity, and delaying transverse acceleration for the first fm/$c$.

It would appear that improving models in all three areas mentioned above is required for rectifying the shortcomings. The default calculation, which includes all three such effects, provides a reasonable description to the data. Without including longitudinal acceleration, which requires a three-dimensional model, it is unreasonable to expect better agreement and it is probably not meaningful to try to better fit the data by adjusting parameters. An additional area of uncertainty documented here comes from the choice of the initial profile, as a more compact source results in a more explosive source. One could reduce any of the three effects mentioned in the previous paragraph, then compensate for them by adjusting the initial density profile.  

A second impression generated by this investigation is that it appears impossible to disentangle the various uncertainties mentioned above by focusing only on two-pion interferometry. Spectra are sensitive to the same model features studied here, as evidenced by the mean $p_t$ values listed in Table \ref{table:meanpt}. Elliptic flow observables, which require a higher-dimensionality model than used here, can also be used to assist in understanding the collision. Hopefully, different observables will be relatively more sensitive to different facets of the model. Then by performing a coordinated analysis of numerous classes of observables, one should be better able to answer specific question and determine specific parameters. These analyses should also incorporate a greater set of femtoscopic measurements, which we list here:
\begin{itemize}
\item Femtoscopy using particles other than pions. Heavier particles are more sensitive to collective flow due to their lower thermal velocity. Correlations between a heavy particle and a light particle, e.g. $p\pi$, are especially sensitive to the patterns of collective flow.
\item Non-central collisions and collisions at different energies. Measurements have already been made as a function of the direction of the pair's momentum relative to the reaction plane \cite{starhbt}. This information is rich in detail, but the meaning of the information is not yet understood. Additionally, there exists data at different beam energies. By studying the response to changes in the initial energy density, without changing the size, one should gain some leverage for disentangling some of the issues mentioned above.
\end{itemize}

Finally, it should be emphasized that, although the large discrepancy with the $R_{\rm out}/R_{\rm side}$ ratio has been eliminated, none of the variations studied here provided a completely satisfactory reproduction of the $p_t$ dependence of source dimensions. The data showed a modest fall of the ratio for higher $p_t$, which combined with the constraint that the ratio must be unity for $p_t=0$, gives a non-monotonic behavior. Although the rise and fall are only of the order of 10\%, the model calculations all showed monotonic behavior with $p_t$. Furthermore, the model calculations tend to over-predict $R_{\rm long}$ at low $p_t$. This might be due to the assumption of boost invariance, which if relaxed, should provide corrections in the direction of the data. Finally, conspicuous by its absence, has been a comparison of the $\lambda$ factors, which represent the fraction of pairs that are correlated. The model calculations over-predicted these factors, but without a better understanding of experimental details about acceptance of weak decay products and particle mis-identification fractions, one can not as yet draw any conclusions.

\section*{Acknowledgments}
Support was provided by the U.S. Department of Energy, Grant No. DE-FG02-03ER41259.

\end{document}